\documentclass[twoside,9pt]{article}
\usepackage{extsizes}
\usepackage[super,sort&compress,comma]{natbib} 
\usepackage[version=3]{mhchem}
\usepackage[left=1.5cm, right=1.5cm, top=1.785cm, bottom=2.0cm]{geometry}
\usepackage{balance}
\usepackage{mathptmx}
\usepackage{amssymb} 
\usepackage{tcolorbox}
\usepackage{etoolbox}
\usepackage{sectsty}
\usepackage{graphicx} 
\usepackage{lastpage}
\usepackage[format=plain,justification=justified,singlelinecheck=false,font={stretch=1.125,small,sf},labelfont=bf,labelsep=space]{caption}
\usepackage{float}
\usepackage{fancyhdr}
\usepackage{fnpos}
\usepackage[english]{babel}
\addto{\captionsenglish}{%
  
}
\usepackage{array}
\usepackage{droidsans}
\usepackage{charter}
\usepackage[T1]{fontenc}
\usepackage[usenames,dvipsnames]{xcolor}
\usepackage{setspace}
\usepackage[compact]{titlesec}
\usepackage{hyperref}

\usepackage{epstopdf}
\definecolor{cream}{RGB}{222,217,201}

\titleformat{\section}          
  {\bfseries\fontsize{12pt}{14pt}\selectfont} 
  {\thesection}                
  {1em}                        
  {}                           
\begin{document}

\begin{flushleft}
{\fontsize{18pt}{20pt}\selectfont\textbf{Mid-infrared photo-induced force microscopy (IR-PiFM/PiF-IR) - Answers to some questions}}
\end{flushleft}
{\fontsize{12pt}{14pt}\selectfont Daniela Täuber\textsuperscript{*a,b}}\\\\

Mid-infrared photo-induced force microscopy (IR-PiFM/PiF-IR) enables high-resolution chemical imaging of surfaces with lateral resolution less than 5~nm. Here are some answers to questions about the physical background, practical handling and potential applications of PiF-IR including its use in the context of studying antimicrobial interaction. Such questions had been addressed to me during the Faraday Discussions on Vibrations at Interfaces which took place in April 2026 in Manchester/UK. The discussion was part of the theme "What is the question, what is the technique?" in the context of which I presented our recent work [James et al., Faraday Discussions, 2026, doi: 10.1039/d6fd00003g]. A modified version of this manuscript will be published in the themed collection "Vibrations at Interfaces" in Faraday Discussions.
\vspace{1cm}

\section{Introduction}
Nanoscale infrared (IR) spectroscopic imaging methods can brigde the gap between high-resolution structural imaging provided by electron microscopy and scanning probe microscopy methods and chemical information provided by conventional infrared spectroscopy using far field illumination and detection either in transmission or in reflection.\cite{shen_scanning_2024} These methods are complementary to surface enhanced and tip enhanced Raman spectroscopy (SERS and TERS) providing access to high-resolution chemical imaging.\cite{xiao_spectroscopic_2018} In general, spatial resolution enhancement is achieved by incorporating plasmonic nanostructures for nearfield enhancement of infrared absorption in the materials. The nanoscale spatial resolution comes along with a high spectral resolution due to powerful IR light sources, such as quantum cascade lasers (QCLs). This enables a nanoscale approach to cutting edge research in the Live and Materials Sciences, for example, by providing access to secondary structure evaluation of proteins,\cite{ahiri_biophysical_2026} which is intimately connected to protein function at multiple levels. The emerging field of approaches can be grouped by their detection schemes in (i) methods employing far-field optical detection, such as surface enhanced infrared absorption spectroscopy (SEIRA),\cite{ataka_biochemical_2007} and the combination of infrared illumination with scattering scanning nearfield optical microscopy (IR s-SNOM and nano-FTIR)\cite{kanevche_infrared_2026} and (ii) the so-called AFM-IR methods,\cite{xie_what_2023} which combine IR illumination with mechanical detection via atomic force microscopy (AFM).

In the context of the Faraday Discussions of Vibrations at Interfaces in Manchester, several studies involving nanoscale infrared spectroscopic imaging techniques have been presented.\cite{davies-jones_unravelling_2026, james_effect_2026, walters_situ_2026, rikanati_ir_2026} Among these approaches, mid-IR photo-induced force microscopy (IR-PiFM/PiF-IR) provides a particular high sensitivity for vibrations at surfaces due to its highly non-linear mechanical force detection based on tip-enhanced IR illumination in AFM. During the discussion on Friday morning related to our Faraday Discussions article,\cite{james_effect_2026} I received and answered several questions about the physical background, practical handling and potential applications of PiF-IR. This discussion will be included and published in a themed discussion article in Faraday Discussions. A modified version is presented in this manuscript, in particular, I re-structured the received questions and my answers aiming at enhancing clarity about method understanding and potential of PiF-IR.

\section{Discussion}
\subsection{What is the typical acquisition time of a full PiF-IR spectrum?}
For setting an appropriate spectral acquisition time in mid-IR photo-induced force microscopy (PiF-IR, IR-PiFM), several parameters have to be considered, such as local chemical heterogeneity in the sample, strength of probed vibrational absorption bands, adjusted (quantum cascade) laser power, and possibly also signal enhancement by gold or silicon substrates. Typical settings for biological cell and tissue samples are $70-100$~s per spectrum covering a spectral range of $800 - 1800$~${\rm cm}^{-1}$ with spectral tuning steps of 1~${\rm cm}^{-1}$. For the perylene monolayer films used in our article,\cite{james_effect_2026} the acquisition times were set to $60-70$~s per spectrum.

\subsection{How long does it take to acquire a 200~nm~x~200~nm PiF-IR contrast image?}
\begin{figure}[h]
\centering
  \includegraphics[height=8.5cm]{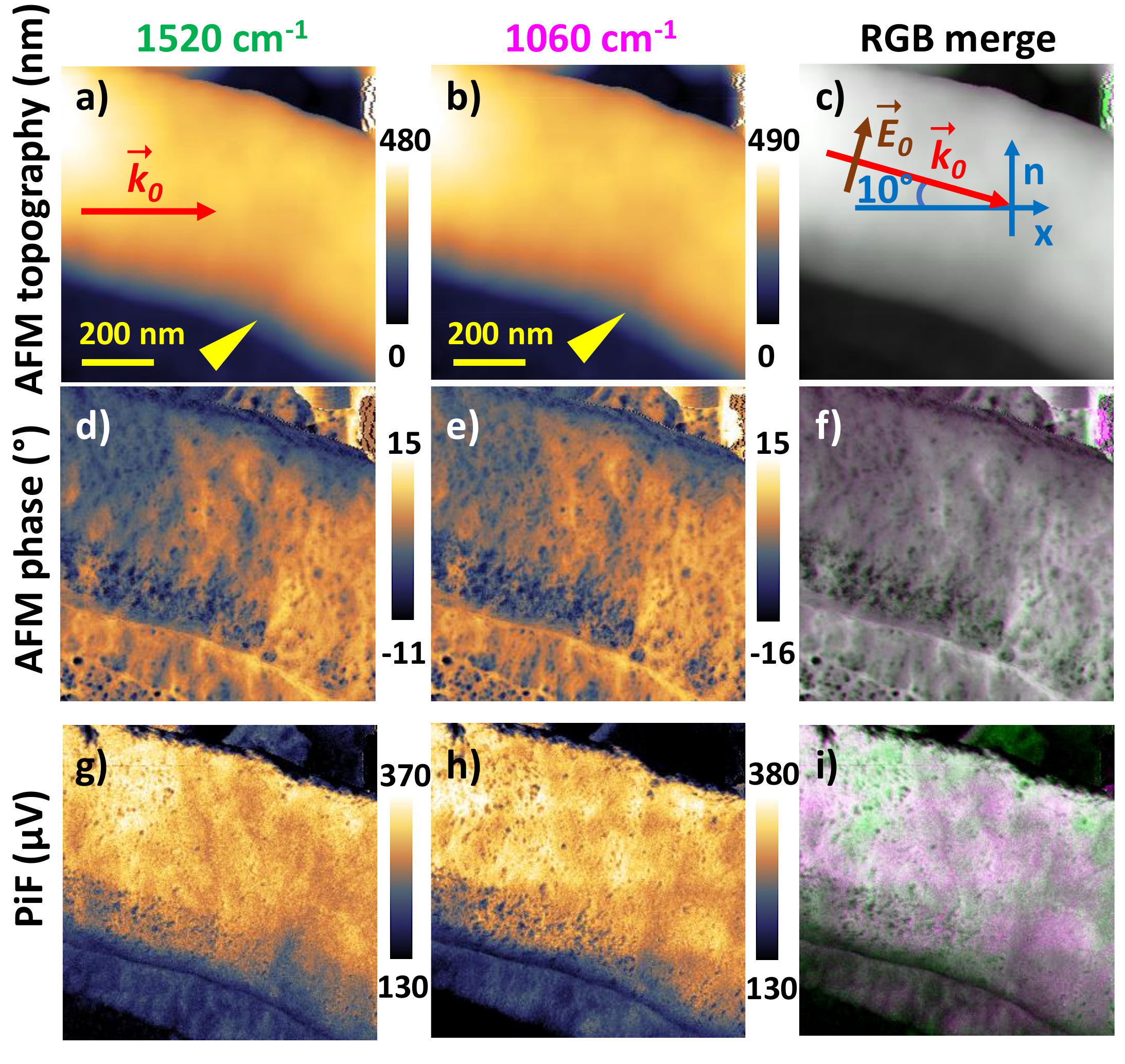}
  \caption{{\bf Single illumination frequency PiF contrasts} of treated \textit{Bacillus subtilis} harvested after 15~min. a-c) Topography, d-f) AFM Phase and g-i) PiF. Left: $\nu=1520$~${\rm cm}^{-1}$, middle: $\nu=1060$~${\rm cm}^{-1}$, right: RGB with "G" set to $\nu=1520$~${\rm cm}^{-1}$ and "R+B" set to $\nu=1060$~${\rm cm}^{-1}$. Schematics in a indicate the light propagation projected onto the sample plane, and in c the illumination geometry including the electric field oscillation in the plane of incidence normal to the sample plane. The yellow arrows in a,b mark a line that is lower in height than the neighboring cell material. Figure reproduced under terms of CC-BY 4.0 license from Ali \textit{et al.}\cite{ali_nanochemical_2025} Copyright ©2025 The Authors. Published by American Chemical Society.}
  \label{fig1}
\end{figure}
The acquisition time of a PiF-IR scan depends on the type, size and pixel-resolution of the scan and on the scan speed. These parameters can be adjusted for imaging particular sample materials and a desired resolution. The 200~nm~x~200~nm scans of the chemically homogeneous and isotropically absorbing polymer nanosphere (NS)\cite{anindo_photothermal_2025} shown in my presentation were acquired with 128 pixels/line and a speed of 0.1 lines/s, resulting in about 20 min per scan and a sampling of approximately 2~nm/pixel. Hyperspectral scans are much slower. Those presented in our paper were acquired over about 14 h each.1
For studying chemical compositions in heterogeneous and nanostructured surfaces, we strongly recommend to study relative intensities using photo-induced force (PiF) contrasts acquired at two or more different infrared (IR) frequencies ($\nu$).1,3
On nanostructured dielectric surfaces, PiF contrasts show anisotropies caused by hybrid field coupling of the infrared illumination with the plasmonic tip and the dielectric sample surface.\cite{james_effect_2026, anindo_photothermal_2025, ali_nanochemical_2025} In our study combining modeling and experiment, we demonstrated that for suitable low illumination power, the shape of the spectra across a scan line is constant, only their overall intensity varies.\cite{anindo_photothermal_2025} Consequently, the chemical composition in heterogeneous surfaces can be deconvolved from anisotropies caused by hybrid field coupling with the nanostructure by evaluating relative PiF intensities acquired using two or more illumination frequencies.\cite{ali_nanochemical_2025} An example is presented in Figure~\ref{fig1} comparing PiF-IR scans of a 1~$\mu$m wide region on the surface of a single bacteria cell acquired using $\nu=1520$~${\rm cm}^{-1}$ (a,d,g) and $\nu=1060$~${\rm cm}^{-1}$ (b,e,h), sensitive for amides and glycans, respectively.\cite{ali_nanochemical_2025} The acquired single-frequency PiF contrasts (g,h) are subject to intensity variations caused by hybrid field coupling of the IR illumination with the nanostructure. A depression line is marked by yellow arrows in the simultaneously acquired topography images (a,b). The right-hand column of the figure shows RGB merge images of the cropped succeeding scans: the topography (c) and AFM phase (f) images confirm a good quality crop, whereas the RGB merge of the PiF (i) reveals a varying chemical composition independent of the intensity variations caused by the field coupling.\cite{ali_nanochemical_2025}
The two PiF-IR scans were sampled at 8~nm/pixels each. To take advantage of the spatial resolution of PiF-IR of less than 5~nm in high-resolution scans, we used 256~pixels/lines for similar 1~$\mu$m-wide areas, resulting in 4~nm/pixels. At a speed of 0.2~$\mu$mm/lines, this results in about 20~min/per scan and a total of roughly 45~min for two successively acquired scans of the same area. Larger overview scans may be acquired faster; however, this may cause contamination of the tip at steep sample edges by overshooting, when the AFM feedback control is not fast enough to maintain the high setpoint in the non-contact dynamic AFM mode employed for PiF-IR.\cite{ali_nanochemical_2025}

\subsection{What is the potential of PiF-IR for providing new insight into biological systems compared to other AFM-IR techniques?}
\subsubsection{So-called AFM-IR methods: detection approaches and probe volumes}
\begin{figure}[h]
\centering
  \includegraphics[height=6cm]{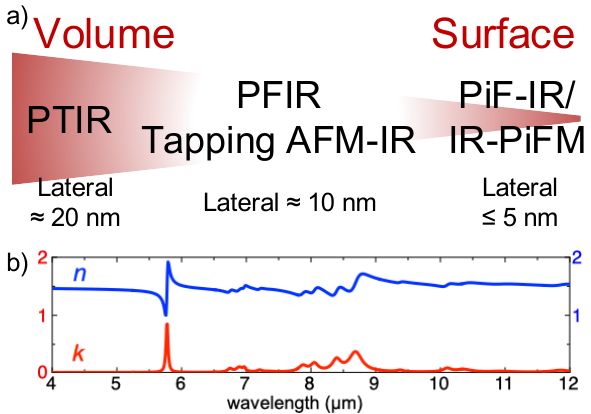}
  \caption{{\bf a) AFM-IR methods} sorted for penetration depths contra surface confinement together with approximate lateral resolutions. b) {\bf Refractive index $n$ and absorption coefficient $k$} of PMMA in the mid-infrared spectral region.}
  \label{fig2}
\end{figure}
The field of high-resolution chemical imaging by combining powerful mid-IR laser illumination with mechanical detection using atomic force microscopy, the so-called AFM-IR methods, has been emerging since the development of photothermal induced resonance (PTIR), which initially coined the term AFM-IR two decades ago.\cite{xie_what_2023} PiFM was initially developed for visible illumination and later was extended to the mid-infrared spectral region. PTIR probes the photo-induced thermal expansion via contact-mode AFM, while PiFM probes the photo-induced force in non-contact dynamic AFM mode, which is achieved by a high set-point for the driven cantilever oscillation in combination with a tiny oscillation amplitude of $1-2$~nm. In contrast, Peak Force Infrared Microscopy (PiFR) and Tapping AFM-IR probe the thermal expansion via intermittent dynamic AFM mode employing oscillation amplitudes in the range of $8-9$~nm.\cite{xie_what_2023} As illustrated in Figure~\ref{fig2}a), the varying AFM-IR approaches result in different penetration depths and lateral resolutions, ranging from volume response in contact mode AFM-IR (PTIR)\cite{jahng_tip-enhanced_2018} to strong surface confinement together with a lateral resolution below 5~nm for IR-PiFM/PiF-IR operated in heterodyne side-band detection mode.\cite{anindo_photothermal_2025, ali_nanochemical_2025, jahng_nanoscale_2019, jahng_quantitative_2022} 

\subsubsection{Physical background of PiFM for visible and mid-IR spectral regions}
As mentioned previously, PiFM was initially developed for visible excitation. In 2014, Jahng \textit{et al.}\ presented a combined theoretical and experimental analysis for this situation.\cite{jahng_gradient_2014} 
In 2016, the group refined the description of the photo-induced force, discriminating different geometries for a sharp metallic tip:\cite{jahng_photo-induced_2016} (i) a tip exposed to an external field in the absence of a substrate, which can be described by the optical tweezer force, (ii) a tip in the presence of a nearby semi-infinite substrate, for which a description considering an image dipole of the tip in the substrate holds, and (iii) a tip in the presence of a small electrically polarizable object, for which they used the description as optical binding force.\cite{jahng_photo-induced_2016} Their theory could successfully describe PiFM spectra in the visible and near-infrared spectral region which involves electronic transitions in the materials. 

Induced-dipole forces follow a dispersive line shape.\cite{jahng_tip-enhanced_2018}
However, in the mid-infrared spectral region, the spectral response of PiFM (IR-PiFM/PiF-IR) usually shows dissipative line shapes (resembling IR absorption spectra), similar to those observed in other AFM-IR techniques.\cite{jahng_tip-enhanced_2018} From this it becomes clear that the measured PiF is modulated by thermal expansion in the tip-sample region. Jahng \textit{et al.} combined modeling and experiment to evaluate the impact of the tip-enhanced thermal expansion force on the behavior of the PiF in the mid-IR spectral region for a metallic tip and a layered planar sample consisting of a dielectric (polymer) and a Si substrate.\cite{jahng_tip-enhanced_2018} In scanning probe microscopy, the tip-sample interaction force $F_{\text{ts}}$ is regarded as the sum of a long-range attractive force and a short-range repulsive force.\cite{bian_scanning_2021}
In the absence of long-range electrostatic forces, the long-range attractive force is the van der Waals (vdW) force interacting between the tip and the sample. This is the general case for PiF-IR probing dielectrics such as polymers and biomaterials.
The viscoelastic tip-sample interaction can be described by various models; a common approach is to use the Derjaguin-Muller-Toporov (DMT) contact force to describe the repulsive force.\cite{jahng_tip-enhanced_2018} 
In general, $F_{\text{ts}}$ depends on the tip-sample distance $z$ and can be expressed as $F_{\text{ts}}(z)=F_\text{c}(z)+F_{\text{nc}}(z)$, 
where $F_\text{c}(z)$ describes the contribution from conservative and 
$F_{\text{nc}}(z)$ non-conservative forces.~\cite{jahng_quantitative_2022} 
For a system consisting of a sharp (spherical) tip and a planar sample,\cite{israelachvili_chapter_2011} the conservative forces are given by~\cite{jahng_quantitative_2022, anindo_photothermal_2025}
\begin{equation*} 
    F_\text{c}(z) \approx 
    \begin{cases}
        \frac{-H_{\text{eff}}r}{12z^2}  & \text{for} \ (z>r_0) \\
        \frac{-H_{\text{eff}}r}{12 r_0^2}+\frac{4}{3}E^*\sqrt{(r_0-z)^3 r} \ \ \ \ &\text{for} \ (z< r_0),
    \end{cases}
\end{equation*}
where $H_{\text{eff}}$ is the effective Hamaker constant that describes the interaction energy between two bodies via a medium, $r$ is the radius of the tip, and $E^*$ is the effective Young's modulus that describes the elasticity of the materials. 
The vdW force depends on the dielectric constants of the materials involved in the interaction.~\cite{israelachvili_chapter_2011} The refractive indices are a macroscopic quantity that describes the interaction of an incident electromagnetic wave with the atoms (electron cloud) of the material and therefore depends on the  density of the material, which, in turn, is decreased upon thermal expansion. If the frequency of the electric field is in resonance with a vibrational transition in the material, it may cause absorption in the material. As a consequence, the refractive index $n(\nu)$ at this frequency $\nu$ diverges, as can be seen exemplarily for the optical constants $n$ and $k$ (absorption coefficient) of polymethylmetacrylate (PMMA) in Fig.\ref{fig2}b. Consequently, the vdW forces are quite sensitive to modulation for spectral frequencies matching absorption bands of the involved materials.
Jahng \textit{et al.}\ showed that for the small cantilever oscillation amplitudes employed in PiF-IR, the heterodyne PiF $F_s$ (side band detection) is proportional to the gradient of $F_{\text{ts}}$:\cite{jahng_quantitative_2022, jahng_nanoscale_2019}
\begin{equation*}\label{eq.thermal force}
    F_s \approx \frac{\delta F_{\text{ts}}}{\delta z}\Delta z.
\end{equation*}
In the heterodyne detection mode, the tip-enhanced thermal expansion is dominated by the non-contact tip-sample interaction force, because the signal vanishes in the contact region, as well as far from the sample, resulting in a high surface sensitivity.~\cite{jahng_nanoscale_2019} The measured PiF in this case is the thermally modulated attractive vdW force.

\subsubsection{Physical background of IR-PiFM/PiF-IR applied to nanostructured dielectric materials}
\begin{figure}[h]
\centering
  \includegraphics[height=6cm]{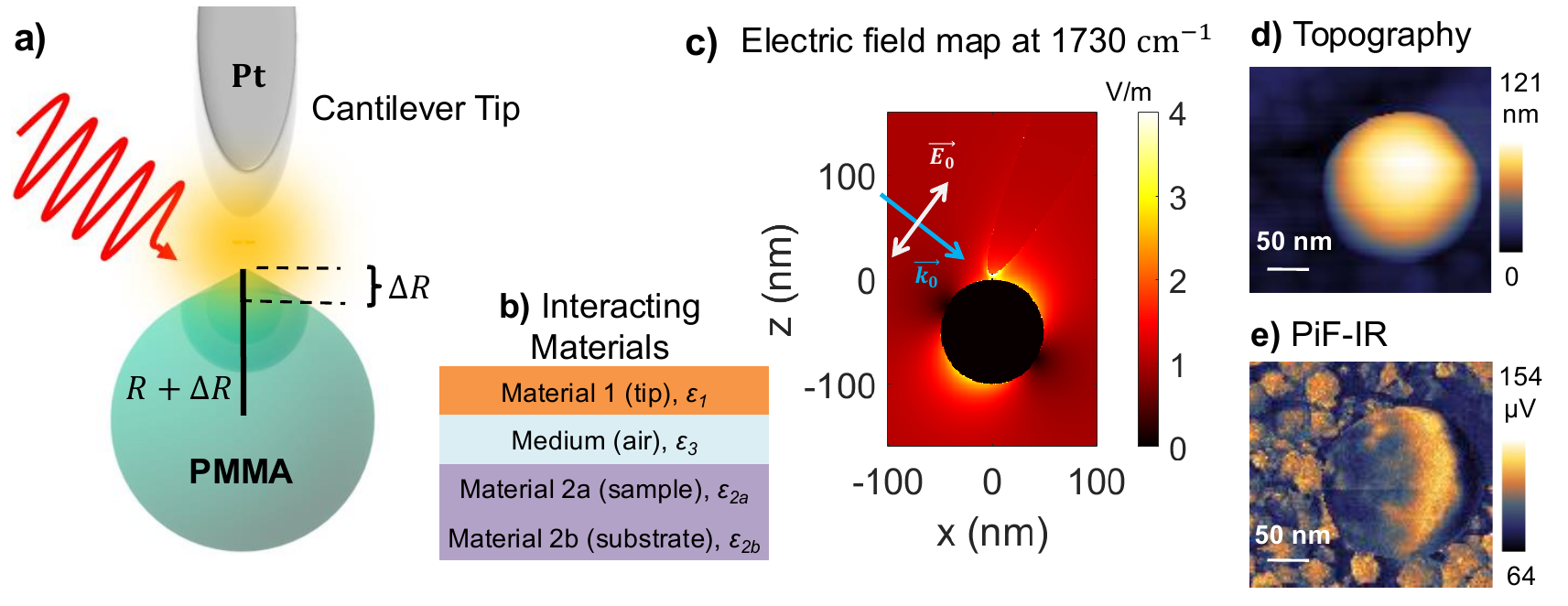}
  \caption{{\bf Hybrid field coupling in PiF-IR.} a) Schematics of photo-induced thermal expansion of a PMMA NS with oscillating Pt tip, b) Material stack involved in van der Waals interaction, c) Modeled electric field for PMMA NS and inclined PT tip for $\nu=1730$~${\rm cm}^{-1}$ and d,e) topography image d and PiF contrast e of experimental PiF-IR scan of PMMA NS. Modified figure reproduced under terms of CC-BY 4.0 license from Anindo \textit{et al.}\cite{anindo_photothermal_2025} Copyright ©2025 The Authors. Published by American Chemical Society.}
  \label{fig3}
\end{figure}
In case of probing nanostructured surfaces using an AFM tip under IR illumination, field coupling of the illuminating electric field with the plasmonic tip and the either plasmonic (metals) or dielectric (soft matter) sample material has to be considered. We combined modeling and experiment to evaluate the effects of photothermal expansion of nanostructured surfaces in photo-induced force microscopy; \cite{anindo_photothermal_2025} the illumination geometry is depicted in Figure~\ref{fig3}a. From this, we found that the hybrid field coupling for a poly(methyl methacrylate) (PMMA) NS in its carbonyl absorption band (1733~${\rm cm}^{-1}$; heat map shown in Figure~\ref{fig3}c) is of the same order of magnitude as the plasmonic field coupling for a gold NS of similar size at the (transversal) surface plasmon resonance of gold (530~nm).\cite{anindo_photothermal_2025} The topography image of the PMMA NS reveals a convolution of its shape with the shape of the probing AFM tip (Figure~\ref{fig3}d), which is typical for scanning probe microscopy.\cite{udpa_deconvolution_2006} The PiF contrast reveals an anisotropic distribution of PiF intensity, which is caused by the hybrid field coupling with the platinum tip and the PMMA NS; see Figure~\ref{fig3}e. Anisotropic intensity distributions obtained on nanostructured dielectric surfaces have been reported in several studies employing various AFM-IR methods.\cite{waeytens_probing_2021, hondl_method_2025, xie_dual-frequency_2022, shcherbakov_photo-induced_2025} 

Evaluating a hyperspectral PiF-IR scan of the PMMA NS, we found that for a suitable low illumination power the shape of the PiF-IR spectrum stayed constant along a scan line across the NS, only its intensity varied.\cite{anindo_photothermal_2025} Consequently, the chemical composition in a surface can be de-convolved from anisotropies introduced by hybrid field coupling by evaluating relative intensities in two or more absorption bands. We applied this in our model study \textit{Bacillus subtils} treated with vancomycin to discriminate areas showing intact bacteria wall from those exposing the underlying cell membrane.\cite{ali_nanochemical_2025} An example is given in Figure~\ref{fig1}.

Another quite important detail for modeling the force between an AFM tip and a nanostructured sample is the dependence of the vdW force on the particular geometry.\cite{parsegian_van_2006, israelachvili_chapter_2011} VdW forces are long-range forces that interact over distances in the range of $10-100$~nm.\cite{israelachvili_chapter_2011, loskill_is_2012,tauber_influence_2013} In practice, this means that in a layered system the subsurface material composition contributes to the effective Hamaker coefficient $H_{\rm eff}$ that describes the interaction; a simple example including a sample on a substrate is depicted in Figure~\ref{fig3}b. In particular, high refractive index materials such as gold and silicon contribute to the long range vdW interaction involving layered systems and nanoparticles (sharp AFM tips) or even single molecules.\cite{loskill_is_2012, tauber_influence_2013} Furthermore, deviations from a planar, layered sample geometry have to be considered separately. The simplified mathematical equations used for describing vdW forces are based on the Derjagun Approximation, that the pairwise contributions of points opposite of each other in two large bodies interacting via a medium can be summed up. The approximation is based on the assumption that cross-interactions approximately cancel due to the overlap of symmetric force contributions.~\cite{israelachvili_chapter11_2011, parsegian_van_2006} Mathematical descriptions of vdW forces for frequently used geometries are listed in Figure~13.1 in Israelachivili, Intermolecular and Surface Forces, 3rd edition, 2011.\cite{israelachvili_chapter_2011} In particular, the apex radius of the tip and the shape of the nanoparticle, for example the radius of a NS, contribute to the vDW force.

\subsubsection{What are the practical limits of PiF-IR in resolving the chemical composition and depth profile of bacterial surfaces?}
\begin{figure}[h]
\centering
  \includegraphics[height=10cm]{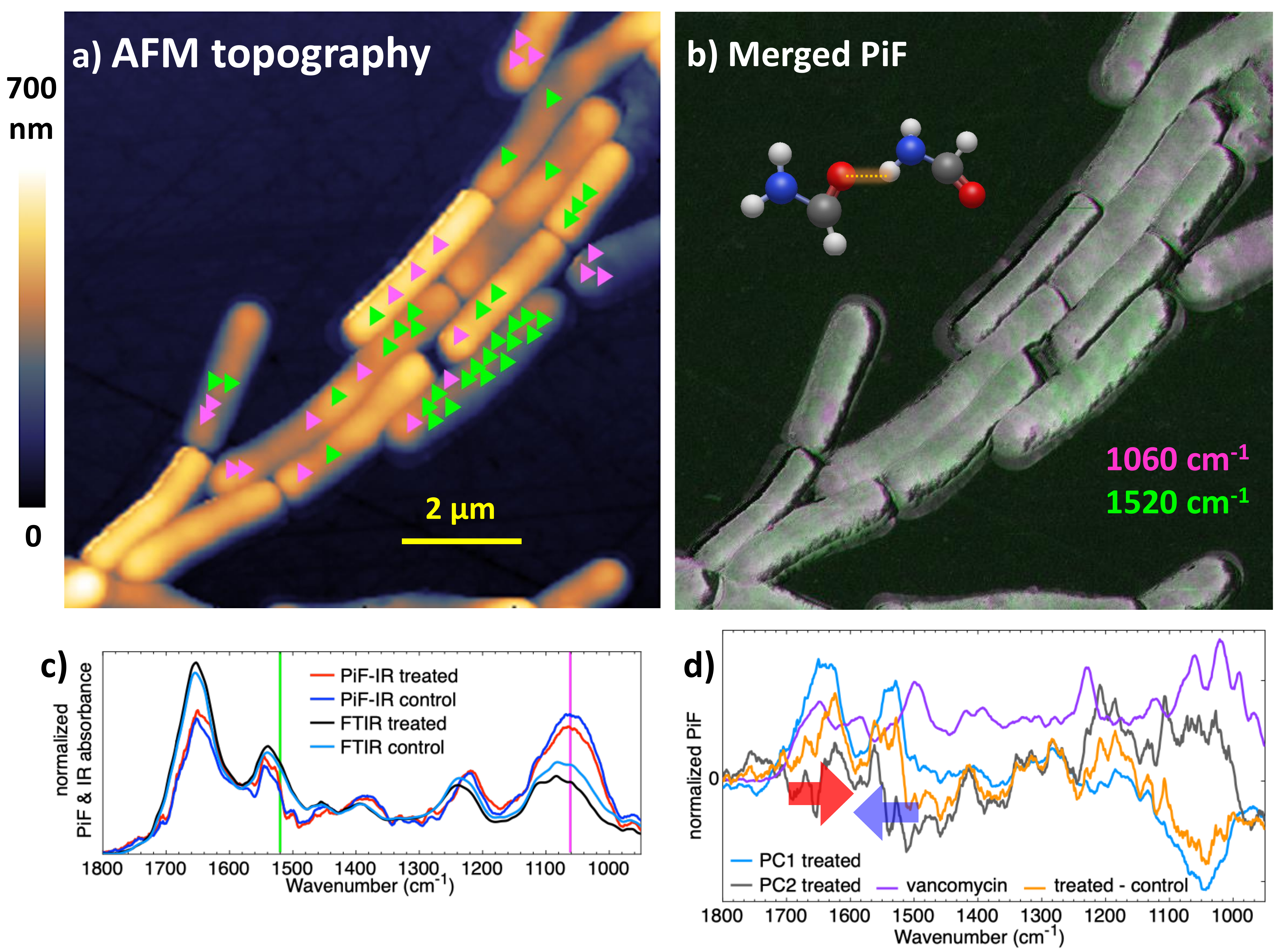}
  \caption{{\bf PiF-IR scans, spectra and chemometrics of treated \textit{Bacillus subtilis} cells harvested after 30~min}. a) AFM topography of  treated cells; positions of point spectra acquisition are marked by triangles following the color code of the “glycan” and “amide” subgroups found by chemometrics analysis; b) Merged PiF contrast from pairs of subsequent sans acquired  at $\nu=1520$~${\rm cm}^{-1}$ (green) and $\nu=1060$~${\rm cm}^{-1}$ (pink); Inset illustrates the hydrogen bond formation between vancomycin and a tripeptide strand in the bacteria wall peptidoglycan; c) mean spectra of PiF-IR spectra acquired on treated (a,b) and control sample and corresponding FTIR spectra; d) difference spectra and chemometrics:  1st and 2nd loadings of PCA applied to PiF-IR spectra from treated cells only, PiF-IR spectrum of vancomycin and difference spectrum of PiF-IR spectra of treated and control presented in c. The red and blue arrows highlight the spectral shifts revealed by PC2 in the amide I and amide II bands, respectively. Modified figure reproduced under terms of CC-BY 4.0 license from Ali \textit{et al.}\cite{ali_nanochemical_2025} Copyright ©2025 The Authors. Published by American Chemical Society.}
  \label{fig4}
\end{figure}
We applied PiF-IR to the investigation of single \textit{Bacillus subtilis} cells treated with vancomycin, a well-studied model system for antibiotic interaction.\cite{ali_nanochemical_2025} Figure~\ref{fig1} shows the chemical contrast on the surface of an untreated control cell acquired at a scan resolution of 8~nm/pixel. In our published study we also presented chemical contrasts of treated \textit{Bacillus subtilis} cells with as high as 4~nm/pixel, which is in the range of the spatial resolution of PiF-IR $\leq 5$~nm. For resolving chemical contrasts of chemically highly heterogeneous surfaces such as mammalian cells and cell organelles, scan resolutions in this range or even better are recommended. Otherwise, the heterogeneous surface of the cell may not be fully sampled. The high stability of the closed-loop scanner of the used VistaScope (Molecular Vista, US) enables several subsequent scans of the same sample position with only small lateral shifts in the range of a few nanometers and almost no scan distortion. In our study of single \textit{Bacillus subtilis} cells, we were interested in the general effect of vancomycin on the bacteria wall, as well as in the localization of the antibiotic interaction.\cite{ali_nanochemical_2025} Figure~\ref{fig4}a shows the AFM topography of a 10~$µ$m-wide overview scan of several \textit{Bacillus subtilis} cells treated with vancomycin for 30~min, which was acquired at 512/pixels per line. The resulting sampling rate of 19.5~nm/pixel does not match the method resolution of less than 5~nm/pixel. Nevertheless, the PiF contrast of the two subsequent scans acquired at $\nu=1520$~${\rm cm}^{-1}$ (green) and $\nu=1060$~${\rm cm}^{-1}$ (pink) shown in Figure~\ref{fig4}b clearly reveals the effect of the antiobiotic interaction on the bacteria wall. $\nu=1060$~${\rm cm}^{-1}$ is sensitive to the peptidoglycan in the bacteria wall, while $\nu=1520$~${\rm cm}^{-1}$ has a higher sensitivity to the underlying cell membrane of these Gram positive bacteria. The PiF contrast shows 50 to several 100 nm wide patches of the exposed cell membrane (green) alternating with areas of a predominantly intact peptidoglycan layer; see Figure~\ref{fig4}b. Vancomycin is known to bind to a tripeptide chain in peptiodoglycan,~\cite{kahne_glycopeptide_2005} hindering cell growth and eventually resulting in cell death.

As mentioned previously, PiF-IR (operated in side-band mode) is surface sensitive. This effect is well seen in the comparison of the PiF-IR spectra acquired from treated and control bacteria with the conventional Fourier Transform Infrared (FTIR) spectra of the same samples;\cite{ali_nanochemical_2025} see Figure~\ref{fig4}c. The normalized comparison reveals a higher intensity of FTIR in the amide I and II bands ($1600-1700$~${\rm cm}^{-1}$ and $1480-1560$~${\rm cm}^{-1}$, respectively) correlated with a lower intensity in the glycan region than the corresponding PiF-IR spectra. FTIR probes the full depth of the bacteria cell, which contains a higher amount of proteins than the surface of the bacteria cells. PiF-IR has a higher sensitivity to the bacteria wall surface, which mainly consists of bacteria wall peptidoglycan. A comparison of PiF-IR spectra with attenuated reflection infrared (ATR) spectra from cancer cell-derived exosomes published by another group further confirms the surface sensitivity of PiF-IR:\cite{kang_nanoscale_2025} Such exosomes are covered by a membrane which has a higher lipid content than their interior. Consequently, the PiF-IR spectra of these exosomes show enhanced intensities in several bands related to lipids compared to the ATR spectra.\cite{kang_nanoscale_2025} The particular penetration depth of PiF-IR depends on the sample composition. It was first demonstrated on an 11~nm thick monolayer of a vertically aligned block-copolymer by Murdick\textit{et al.}\ in 2017.\cite{murdick_photoinduced_2017} 

\subsubsection{New insight into biological systems provided by nano-infrared (IR) spectroscopic imaging methods}
As detailed above, PiF-IR provides an unprecedented spatial resolution of less than 5~nm of the surface of cells.\cite{ali_nanochemical_2025} It is less sensitive to volume response in cells than the other AFM-IR methods.
Thus, access to complementary new information about biological systems can be provided by a
combination of different nano-IR spectroscopic methods, including PiF-IR,\cite{davies-jones_unravelling_2026, ali_nanochemical_2025, shcherbakov_photo-induced_2025} other AFM-IR methods,\cite{ruggeri_infrared_2021} and methods using far-field optical detection of nearfield-enhanced IR absorption in biological systems, such as mid-IR s-SNOM/nano-FTIR\cite{kanevche_infrared_2026} and SEIRA.\cite{ataka_biochemical_2007} 

IR spectroscopy is widely used to probe not only the chemical composition of biological systems but also to provide details about molecular environments. Several nano-IR-spectroscopic imaging methods, including PiF-IR, have been used to study the secondary structure of proteins,\cite{joseph_nanoscale_2024, ruggeri_infrared_2021, ahiri_biophysical_2026} which is closely related to the functionality of proteins in biological systems. The combination of a high spatial resolution with a high spectral resolution in PiF-IR provides access to highly localized spectral information on chemically heterogeneous surfaces. Details about local molecular orientation and molecular environment are hidden by large-area averaging in conventional IR spectroscopic imaging methods, such as FTIR and ATR.\cite{ali_nanochemical_2025} To fully exploit the potential of PiF-IR and other nano-IR spectroscopic imaging methods, IR spectra of biological systems and molecular components must be known in more detail than in conventional IR spectroscopy. For this reason, I have started an initiative for an open source data infrastructure which can help to compare new results with published spectra of related components from biological systems. 

\subsection{Can PiF-IR provide meaningful insight into the efficacy of new and alternative antimicrobial strategies?}
We used \textit{Bacillus subtilis} and vancomycin, a well-studied model system for antibiotic interaction, to demonstrate the potential of PiF-IR/IR-PiFM to provide chemical contrasts of the surface of treated cells with a spatial resolution of less than 5~nm.\cite{ali_nanochemical_2025} PiF-IR can reveal highly localized information on the response of individual bacteria cells to antibiotic treatment. An example showing several treated \textit{Bacillus subtilis} is presented in Figure~\ref{fig4}. A chemometrics analysis revealed further details of the interaction. The major chemical variation (first principal component: PC1) in a set of 50 single point PiF-IR spectra acquired in this sample discriminated two subsets: (i) spectra reporting the presence of peptidoglycan and (ii) spectra revealing exposure of the underlying cell membrane; see PC1 (blue line) in Figure~\ref{fig4}d. The positions of the single point spectra were colored according to their PC1 scores in the topography image Figure~\ref{fig4}a; showing peptidogylcan (pink, positive PC1 score) and exposure of cell wall (green, negative PC1). In general, the color code agrees with the color in the merged PiF contrast of the two high-resolution PiF-IR scans acquired subsequently at spectral frequencies sensitive for the peptidoglycan layer and for the underlying cell membrane, at $\nu=1060$~${\rm cm}^{-1}$ (pink) and $\nu=1520$~${\rm cm}^{-1}$ (green), respectively;\cite{ali_nanochemical_2025} see Figure~\ref{fig4}b. The second principal component (PC2) reports chemical shifts in the spectra, which can be assigned to the formation of hydrogen bonds between vancomycin and a tripeptide chain of peptidoglycan.\cite{kahne_glycopeptide_2005, ali_nanochemical_2025} The observed redshift in the amide I band is correlated with a blueshift in the amide II band, which is characteristic for hydrogen bond formation between amide groups.\cite{barth_infrared_2007, ali_nanochemical_2025}
We used these results to locate the binding of vancomycin to peptidoglycan in two hyperspectra acquired from the surface of individual \textit{Bacillus subtilis} cells treated with vancomycin for 30 and for 60~min.\cite{ali_nanochemical_2025} 
This example shows the potential of PiF-IR to reveal local chemical variations in the surface of bacteria cells treated with an antibiotic drug. Furthermore, PiF-IR was successfully applied to histological tissue in the context of polymicrobial infection\cite{joseph_nanoscale_2024} and to study the surface of individual mammalian cells.\cite{davies-jones_unravelling_2026} These studies demonstrate further possibilities of using PiF-IR in the context of antibiotic discovery to investigate the response of mammalian cells and tissue to infection and antibiotic treatment.

PiF-IR measurements must be carefully evaluated for possible contamination, which even in non-contact AFM methods such as PiF-IR can be introduced by the so-called overshooting at steep edges in the sample.\cite{ali_nanochemical_2025} Such contamination is even more likely to occur in contact mode or intermittent mode AFM-IR methods. PiF-IR provides access to evaluation of such contamination because the method is highly sensitive to even tiny amounts of material, which may contribute to results using other AFM-IR techniques as well with less access for its detection.

\subsection{How can electric interactions in samples of unknown chemical compositions like cells be modeled?}
The situation is even more complicated because even for known chemical compositions, the refractive indices of biomaterials in the mid-infrared spectral region are usually not really known. A typical approach is to use known refractive indices of polymers for modeling.\cite{anindo_photothermal_2025, tsuda_spectral_2018, pascual_robledo_theoretical_2025} Apart from the specific absorption bands, the refractive indices of polymer materials and many organic materials do not differ much.

\subsection{Can PiF-IR be used for imaging live cells, to do real-time imaging and to distinguish Gram positive and negative bacteria?}
PiF-IR probes tip-sample interaction forces of the order of nN.\cite{jahng_nanoscale_2019, sifat_photo-induced_2022} Live cell imaging usually requires a liquid environment, which introduces substantial damping to the cantilever oscillation. From this perspective, it is quite unlikely that it will be possible to extend PiF-IR to measurements in liquids. However, the Davies group in Cardiff has successfully conducted PiF-IR measurements of living cells using a chamber that provides constant humidity at 70\%.\cite{davies-jones_unravelling_2026} Real-time imaging is another challenge for scanning probe microscopy. So far, typical PiF-IR scans require acquisition times in the range of several minutes to hours, depending on the sampling frequency and the spectral range. Under these circumstances, applications for real-time imaging are rather unlikely, although perhaps possible for tailored experimental designs. Given the power of PiF-IR for a high spatial resolution of surfaces\cite{ali_nanochemical_2025} and the distinct difference in the bacteria wall of Gram positive and Gram negative bacteria, it is, of course, possible to distinguish them, as recently demonstrated by Davies-Jones \textit{et al.}\cite{davies-jones_photoinduced_2024}

\subsection{Can the interaction with the AFM tip cause local changes in osmotic pressure or solvation structure and how might this affect the measurement?}
So far, PiF-IR measurements have not been conducted in solutions, which would be a pre-requisite for studying the influence of solvation effects. Under ambient and dry conditions, hydrophilic surfaces are covered by a water layer that varies in thickness with variation of the relative humidity. In ambient AFM, such water layers are known to affect the measured signal. Metallic AFM tips are hydrophilic and thus are usually covered by a water layer under ambient conditions. So far, there have been only very few applications of PiF-IR/IR-PiFM to biological cells and other histological samples\cite{shcherbakov_photo-induced_2025, davies-jones_photoinduced_2024, ali_nanochemical_2025, kang_nanoscale_2025} and the effects of the surface water layer have not been investigated.

\subsection{Can functionalization of the tip chemistry improve biocompatibility while preserving the optical enhancement?}
Spectral PiF intensities are sensitive to molecular contamination at the tip.~\cite{james_effect_2026} In particular, polydimethylsilane (PDMS) is frequently present in scanning probe microscopy labs, as it is used to store AFM tips. Jahng \textit{et al.}\ studied the effect of a tip functionalized by PDMS on signal detection and found that the thermal expansion of PDMS could enhance the contrast of polyacrylonitrile-nanocrystalline cellulose nanofibers on a glass surface due to matching spectral bands.\cite{jahng_substructure_2018} Biological systems exhibit considerably higher spectral complexity than isolated and chemically modified cellulose fibers on glass surfaces. Thus, it will be much more complex to find and control a suitable tip functionalization for studying such systems.

\subsection{Does PiF-IR reach single-molecule sensitivity?}
In our experimental approach employing sharp platinum-coated AFM tips, PiF-IR probes a volume with about 5 nm lateral extension and 5-10 nm depth. Consequently, the answer to this question depends on the size of the probed molecules. The rather rigid aromatic core of the alkylated perylene monoimides investigated in our article by James et al.\cite{james_effect_2026} has a length of a few nanometers, but a smaller width and height. It is known to form dimers on silver substrates.\cite{hupfer_supramolecular_2021} In this case, the number of probed molecules can be estimated between 2 and a few, depending on the molecular orientation in respect to the surface of the gold substrate.
The other spectra I presented during the discussion were obtained from our model study of antibiotic interaction: bacillus subtilis and vancomycin.\cite{ali_nanochemical_2025} The bacteria wall consists mainly of the biopolymer peptidoglycan, which forms a micrometer-sized network of fibers. Vancomycin is a rather flexible glycopeptide\cite{kahne_glycopeptide_2005} of a few nanometers in size. We used the spectral signature of hydrogen bond formation obtained from a chemometrics analysis of individual PiF-IR spectra (see Figure~\ref{fig4}) to locate the characteristic hydrogen bond formation in hyperspectral scans of the surface of bacteria with a sampling of approximately 10~nm/pixel. A higher sampling rate that meets the spatial resolution of PiF-IR may come closer to single-molecule resolution. However, it is quite difficult to estimate the molecular sensitivity in a complex biological sample, in particular, without the proximity of a high-refractive-index substrate. 

\subsection{Can single-molecule sensitivity be reached by diluting the sample?}
In principle, sensing single molecules using PiF-IR can be possible in carefully tailored experiments, for example on gold substrates. High refractive index substrate materials can improve the sensitivity of PiF-IR by enhancing the van der Waals interaction between the tip and the sample material. Typically used \ce{CaF2} substrates are not suited for single-molecule detection, in such substrates, the weak PiF-IR signal will be drowned in the instrument noise.

\subsection{Kanehira et al. used DNA origami to orient molecules and probe them using SERS. Would it be interesting to apply PiF-IR to molecules with modified oriententations provided by DNA origami?}
Thank you for pointing out this interesting work by Kanehira \textit{et al.}\cite{kanehira_molecular_2026} The application of PiF-IR to a selection of IR-active single molecules aligned on top of a DNA strand could provide a suitable test for the potential single molecule sensitivity of the method. For achieving such a sensitivity, it will be important to enhance the attractive vdW force acting in the tip-sample region by use of high refractive index substrates, preferably gold. In this case, the electric fields are known to be oriented perpendicular to the substrate.\cite{james_effect_2026, pascual_robledo_theoretical_2025} Therefore, I do not expect to observe a difference between vibrational transition moments aligned parallel or perpendicular to the DNA-strand within the sample plane. However, orientations parallel and perpendicular to the substrate can be discriminated.\cite{james_effect_2026} My own access to PiF-IR instrumentation has been quite limited, which prevents me from conducting such tests in my group. Yet, there might be other groups that have access to an instrument and who are interested in joining a collaborative study. 

It might be an idea to complement a  possible PiF-IR investigation by surface enhanced IR absorption spectroscopy (SEIRA) for comparison of vibrational bands and their orientations. For SEIRA, suitable plasmonic enhancement can be achieved by employing the longitudinal mode of surface plasmons of rod-shaped gold nanoantennas,\cite{milekhin_nanoantenna_2018} which can be tuned to the mid-infrared by adjusting the rod length. 

\subsection{Does the choice of substrate material affect the shape of PiF-IR, in particular, by affecting contributions from in-plane and out-of-plane vibrational modes?}
PiF-IR operated in the heterodyne detection scheme is highly surface sensitive due to the highly non-linear dependence of the measured PiF on IR absorption in the sample material.\cite{jahng_nanoscale_2019, jahng_quantitative_2022} We conducted a study on layered PMMA films comparing experimental PiF-IR spectra with experimental and calculated FTIR spectra. Our unpublished results showed variations in the PiF-IR spectral shape with increasing thickness $d$ of the PMMA film until a certain threshold for $d$, in agreement with observations by Jahng \textit{et al.}.\cite{jahng_nanoscale_2019} In contrast to FTIR,\cite{mayerhofer_removing_2018, mayerhofer_caf_2020, mayerhofer_removing_2020} the recorded PiF-IR spectra did not show a considerable dependence on substrate materials. However, we did not study films with $d<60$~nm. For monolayer films of an alkylated perylene-monoimide on gold substrates, we observed a strong dependence of orientation-sensitive vibrational transitions on their orientation relative to the surface normal on the gold substrate.\cite{james_effect_2026} This agrees with a recent study by the Hillenbrand group who used theoretical modeling to study in-plane and out-of-plane vibrations for layered materials on strongly and weakly reflecting substrates.\cite{pascual_robledo_theoretical_2025}  An experimental study applying PiF-IR to molecular monolayers on weakly reflecting substrates such as \ce{CaF2} is not feasible due to the much weaker vdW forces acting between the tip and the layered sample in that case. The contribution of the substrate to the long-range vdW force in the tip-sample region decreases with increasing distance, which leads to the previously described saturation effect for changes in spectral shapes with increasing film thickness.\cite{jahng_nanoscale_2019} Thus, for large enough sample thickness, the field enhancement is only influenced by hybrid field coupling with the nanostructure in the tip-sample region.\cite{anindo_photothermal_2025} From modeling fields in the vicinity of a sharp metallic tip, it is known, that these fields also contains components parallel to the substrate.\cite{pascual_robledo_theoretical_2025, waeytens_probing_2021} Further experimental work and modeling is required to improve the understanding of such effects for possible impacts on PiF-IR spectra.

\footnotetext{\textit{$^{*}$~E-mail: dantaube@gmx.de}}
\footnotetext{\textit{$^{a}$~Institute of Physical Chemistry (IPC), Friedrich Schiller University Jena, 07743 Jena, Germany.}}
\footnotetext{\textit{$^{b}$~Leibniz Institute of Photonic Technology (LIPHT), 07745 Jena, Germany.}}

\section*{Conflicts of interest}
There are no conflicts to declare.


\section*{Acknowledgements}
I thank Philip R. Davies for organizing and chairing the Faraday Discussions on Vibrations at Interfaces and Renee Frontiera for chairing the Friday morning session including the discussion which provided the grounds for this manuscript. Many thanks also to Zsuzsanna Heiner, Jacob Pattem, Eric Borguet, Renee Frontiera, Rohit Chikkaraddy and Sergio Kogikoski Jr.~for their posed questions about PiF-IR and its possible applications.
For the VistaScope and the MirCat QCL financial support of the European Union via the Europäischer Fonds für Regionale Entwicklung (EFRE) and the Thüringer Ministerium für Wirtschaft, Wissenschaft und Digitale Gesellschaft (TMWWDG) is acknowledged (Projects: 2018 FGI 0023 and 2023 FGI 0018). My work using PiF-IR has been supported by a postdoctoral Scholarship (PolIRim) from Friedrich Schiller University Jena in 2020 and by a project grant (pintXsum) together with Christin David in the context of the profile lines \textit{Light} and \textit{Life} in 2022/23. I further acknowledge funding from the German Research Foundation (DFG, Projects: 542825796 "HiResi4RPE" and 439139881 "Live2DPOLIM"). Many thanks also to Rainer Heintzmann, who has been hosting my BioPOLIM group (https://biopolim.de/) in his Bio-Nanoimaging Group (https://nanoimaging.de/) at the Friedrich Schiller University Jena and the Leibniz Institute for Photonic Technology in Jena since the end of 2019.







\bibliography{AnswersPIFIR} 
\bibliographystyle{rsc} 

\end{document}